# Multilevel Clustering Fault Model for IC Manufacture

Yu. I. Bogdanov [*], N. A. Bogdanova[**], A. V. Rudnev[*]

[*]OAO Angstrem, 124460 Moscow, Russia[1]
[**]Moscow Institute of Electronic Engineering (Technical University), 124460 Moscow, Russia

**ABSTRACT**

A hierarchical approach to the construction of compound distributions for process-induced faults in IC manufacture is proposed. Within this framework, the negative binomial distribution is treated as level-1 models. The hierarchical approach to fault distribution offers an integrated picture of how fault density varies from region to region within a wafer, from wafer to wafer within a batch, and so on. A theory of compound-distribution hierarchies is developed by means of generating functions. A study of correlations, which naturally appears in microelectronics due to the batch character of IC manufacture, is proposed. Taking these correlations into account is of significant importance for developing procedures for statistical quality control in IC manufacture. With respect to applications, hierarchies of yield means and yield probability-density functions are considered.

## 1. INTRODUCTION

As early as in 1964, Murphy [1] introduced the compound Poisson distribution into microelectronics, having found that the traditional Poisson distribution is not always adequate to predict yield in integrated-circuit (IC) manufacture. The point is that process-induced faults tend to occur unevenly over the wafer, appearing as clusters. Murphy's approach was developed by Seeds [2], Okabe et al. [3], Stapper [4, 5], and other researchers.

The compound Poisson distribution differs from the traditional one in that the parameter $\lambda$ denoting the fault density is regarded as a random variable. Experience indicates that the gamma distribution is probably the most accurate model for $\lambda$ [6--10]. Also note that the compound Poisson distribution is a limiting case of the compound binomial distribution, the latter arising within Polya's urn model [11-12].

Statistical yield models based on compound distributions have proven to be useful in design, manufacture, and product evaluation alike. Yield enhancement aims to make the product fault-tolerant (i.e., less sensitive to process-induced faults) by adding a degree of redundancy to the IC (error-correcting codes are an example) and by optimizing its floorplan and layout [13--18]. Concerning the manufacture phase, we cite the Bayesian approach. Applied to in-process product control, the method allows one to refine yield prediction and make decisions as the batches progress along the processing line [19]. Compound-distribution models also help one to calculate yield distribution over wafers, to estimate costs, to evaluate manufacturing efficiency, to predict yield losses, etc. [20, 21].

In this paper, we describe a hierarchical approach to the construction of compound distributions for process-induced faults in IC manufacture.

Section 2 describes the origin and main properties of a yield model that is built around the compound Poisson distribution and has been accepted by the electronics industry [17].

Section 3 develops a theory of compound-distribution hierarchies. Within this framework, the Poisson distribution and the negative binomial distribution belong to level 0 and 1. It is shown that main formulae can be written in compact analytical form by means of generating functions.

Section 4 deals with applied aspects. Compared with previous results, a more general formalism for mean yields is presented. Also included are equations for yield probability densities. The yield is thus treated as a hierarchical random variable.

## 2. COMPOUND POISSON DISTRIBUTION

Primitive yield models employ the binomial distribution. Consider a chip with $n$ components. If each of them has the same probability $p$ of being faulty and faults arise independently of one another, then the probability $P(k)$ that the chip has exactly $k$ faulty components is

$$P(k) = C_n^k \cdot p^k \cdot (1-p)^{n-k}, \tag{1}$$

where
$k = 0, 1, ..., n$, $0 < p < 1$.

Since $n$ is very large and $p$ is very small, the binomial distribution can be approximated by a Poisson distribution to a good accuracy. Accordingly, with $\lambda$ denoting $np$ (the expected value of $k$), the probability $P(k)$ is given by

$$P(k) = \frac{\lambda^k}{k!} e^{-\lambda} \qquad k = 0, 1, 2, ... \tag{2}$$

Assume that the chips have no redundancy, so that any faulty component will make the chip nonconforming. Then, the yield $Y$ is equal to the probability that a randomly chosen chip is fault-free:

$$Y = P(k=0) = e^{-\lambda}. \tag{3}$$

However, it was found as early as in 1964 that Eq. (3) makes badly pessimistic predictions if applied to large-area chips [1]. The underlying reason, as was revealed later, is that process-induced faults do not arise independently in different regions of the wafer but tend to cluster.

To allow for fault clustering, the compound Poisson distribution was introduced, in which the expected number of faults per chip, $\lambda$, is also a random variable. Let $P(\lambda)$ be the probability-density function (PDF) of $\lambda$. The compound Poisson distribution is defined as

$$P(k) = \int_0^\infty \frac{\lambda^k}{k!} e^{-\lambda} P(\lambda) d\lambda, \tag{4}$$

so that

$$Y = P(k=0) = \int_0^\infty e^{-\lambda} P(\lambda) d\lambda. \tag{5}$$

The density $P(k)$ might be specified in various forms. Murphy [1] proposed the triangular distribution

$$P(\lambda) = \begin{cases} \dfrac{\lambda}{\lambda_0^2} & 0 \leq \lambda \leq \lambda_0 \\ \dfrac{2\lambda_0 - \lambda}{\lambda_0^2} & \lambda_0 \leq \lambda \leq 2\lambda_0 \\ 0 & \lambda > 2\lambda_0 \end{cases}, \tag{6}$$

where $\lambda_0$ is the average number of faults per chip. Equations (5) and (6) imply that

$$Y = \int_0^\infty e^{-\lambda} P(\lambda) d\lambda = \left(\frac{1 - e^{-\lambda_0}}{\lambda_0}\right)^2. \tag{7}$$

Seeds [2] assumed that

$$P(\lambda) = \frac{e^{-\lambda/\lambda_0}}{\lambda_0}, \tag{8}$$

so that

$$Y = \int_0^\infty e^{-\lambda} P(\lambda) d\lambda = \frac{1}{1+\lambda_0}. \tag{9}$$

Okabe et al. [3] and Stapper [4, 5] specified $P(\lambda)$ as a gamma distribution:

$$P(\lambda) = \frac{b^a \lambda^{a-1} e^{-b\lambda}}{\Gamma(a)}, \tag{10}$$

where $a$ and $b$ are positive parameters and $\Gamma(a)$ is the gamma function [22, 23]. This distribution has the mean $a/b$ and the variance $a/b^2$.

Equation (4) thus becomes

$$P(k) = \int_0^\infty \frac{\lambda^k}{k!} e^{-\lambda} P(\lambda) d\lambda = \frac{\Gamma(k+a) b^a}{k! \Gamma(a)(1+b)^{k+a}}. \tag{11}$$

Distribution (11) is commonly known as the negative binomial distribution. With two parameters available for adjustment, one can fit the model to observation data in terms of variance as well as mean.

In what follows the density $P(\lambda)$ appearing in Eq. (4) will be taken in form (10). This type of compound Poisson distribution allows one to accurately represent real situations and has some theoretical advantages (see below).

The mean and variance are given by

$$\mu = \frac{a}{b}, \tag{12}$$

$$\sigma^2 = \frac{a}{b^2}(1+b). \tag{13}$$

It is seen from Eqs. (12) and (13) that $\sigma^2 > \mu$, whereas the traditional Poisson distribution has equal mean and variance.

Solving Eqs. (12) and (13) for $a$ and $b$, we obtain

$$a = \frac{\mu^2}{\sigma^2 - \mu}, \tag{14}$$

$$b = \frac{\mu}{\sigma^2 - \mu}. \tag{15}$$

Formulas (14) and (15) are useful for fitting the distribution to a sample, with the sample mean and variance assigned to $\mu$ and $\sigma^2$, respectively.

Two limiting cases are worth noting: (i) If $a$ and $b$ tend to infinity in such a way that $a/b$ tends to a finite number $a/b \to \lambda_0 = const$, then the compound Poisson distribution approaches the traditional Poisson distribution in which $\lambda = \lambda_0$. (ii) If $a$ tends to infinity and $b$ is fixed, then the compound Poisson distribution approaches the normal distribution for which Eqs. (12) and (13) hold.

Equation (11) implies that

$$Y = P(k=0) = \frac{1}{\left(1 + \frac{1}{b}\right)^a}. \tag{16}$$

Let us recast this in terms of the average number of faulty components per chip, $\lambda_0 = \frac{a}{b}$:

$$Y = P(k=0) = \frac{1}{\left(1 + \frac{\lambda_0}{a}\right)^a}. \tag{17}$$

The model considered is known as the large-area clustering negative binomial model [13, 15]. The parameter $a$ is called the cluster parameter. Its typical values approximately range from 0.3 to 7. In actual fact, fault clustering disappears if $a$ exceeds 4 or 5; for such $a$, Eq. (17) can be approximated by formula (3).

The large-area clustering model is based on two assumptions. First, fault clusters are larger than chips, so that any faulty chip is totally covered by one fault cluster. Second, faults are distributed uniformly within any cluster. In addition, there are the small-area clustering negative binomial model [24] and the medium-area clustering negative binomial model [25, 26]. The latter is regarded as including the other models [13]. It is intended for chips with areas on the order of a square inch; they may well be larger than fault clusters. Since the concept of fault cluster has yet to be clarified, the medium-area clustering model is defined in terms of blocks [25]. It is assumed that (i) correlation between faults may exist only within a block, (ii) blocks are statistically independent of each other, (iii) the total number of faults per block obeys a negative binomial distribution, and (iv) faults are distributed uniformly over each block.

## 3 COMPOUND-DISTRIBUTION HIERARCHIES

### 3.1 Generating-Function Hierarchy for a Compound Poisson Disiribution

Let us describe the multilevel clustering fault hierarchy. At level 0, we have a binomial distribution. It is specified by a single parameter, $p_0$, which might be viewed as the proportion of faulty components. In passing to level 1, we declare the parameter $p_0$ a random variable with a beta distribution and eliminate $p_0$ by the compounding procedure. As a result, we deal with parameters $a_1$ and $b_1$ instead of $p_0$. Finally, we introduce clustering factor

$$g_1 = \frac{1}{a_1 + b_1} \tag{18}$$

and probability of being faulty

$$p_1 = \frac{a_1}{a_1 + b_1}. \tag{19}$$

In general, level $r+1$ is constructed from level $r$ by treating the parameter $p_r$ as a random variable obeying the beta distribution with parameters $a_{r+1}$ and $b_{r+1}$; when passing to level $r+1$, the variable $p_r$ is replaced with the parameters $p_{r+1}$ and $g_{r+1}$ by the compounding procedure. For each $r$,

$$g_r = \frac{1}{a_r + b_r} \tag{20}$$

and

$$p_r = \frac{a_r}{a_r + b_r}. \tag{21}$$

For each level $r$ of the hierarchy, a compound Poisson distribution arises if $p_r \to 0$, $g_r \to 0$, and $n \to \infty$ in such a way that $np_r \to \lambda_r = const$ and $ng_r \to \lambda_r / a_r = const$, where $\lambda_r$ and $\lambda_r / a_r$ are finite numbers [12]. In the microelectronics context, $n \geq 10^6$ and $p_r \leq 10^{-6}$, which refer to IC complexity and the probability of an individual IC component being faulty, respectively. Thus, the above conditions are fulfilled.

At level 0, the binomial generating function changes into the Poisson generating function:

$$G_0(z|p_0, n) = (1 - p_0(1-z))^n = \left(1 - \frac{np_0}{n}(1-z)\right)^n \to \exp(-np_0(1-z)). \tag{22}$$

For levels 0--2, the natural logarithms of the generating functions are given by

$$\ln G_0(z|n, p_0) = -np_0(1-z), \tag{23}$$

$$\ln G_1(z|n, g_1, p_1) = -\frac{p_1 n}{g_1 n} \ln(1 + g_1 n(1-z)), \tag{24}$$

$$\ln G_2(z|n, g_1, g_2, p_2) = -\frac{p_2 n}{g_2 n} \ln\left(1 + \frac{g_2 n}{g_1 n} \ln(1 + g_1 n(1-z))\right). \tag{25}$$

The factors $n$ are retained in fractions in order to show that indefinitely small and large quantities appear as products only.

In general,

$$\ln G_r = -p_r n \cdot L_r. \tag{26}$$

where

$$L_{r+1} = \frac{\ln(1 + g_{r+1} n L_r)}{g_{r+1} n}, \quad L_0 = (1-z). \tag{27}$$

The clustering factor $g_r$ varies with $r$. Concerning the probability $p_r$, it is associated with the highest level and is evaluated by averaging over all the levels. Accordingly, one could simply write $p$ instead of $p_r$.

### 3.2 PMF and Moments

A generating function contains the complete information on the random variable. Consider a discrete random variable with the generating function $G(z)$. The probability that the random variable takes a value $k$ can be expressed in terms of the $k$ th derivative of $G(z)$ at $z = 0$, and an $m$ th factorial moment is equal to the $m$ th derivative of $G(z)$ at $z = 1$:

$$P(k) = \frac{1}{k!} \frac{\partial^k G(z)}{\partial z^k}\bigg|_{z=0}, \tag{28}$$

$$E[k(k-1)...(k-m+1)] = \frac{\partial^m G(z)}{\partial z^m}\bigg|_{z=1}. \tag{29}$$

In particular, the expected value and variance of a discrete random variable $k$ are given by

$$E[k] = G'(1), \tag{30}$$

$$Var[k] = G''(1) + G'(1) - G'^2(1). \tag{31}$$

The above formulae imply that the mean and variance for the compound distribution of level $r$ can be expressed in simple form:

$$\mu = np, \qquad (32)$$

$$\sigma^2 = np(1 + g_1 n + g_2 n + \ldots + g_r n). \qquad (33)$$

In Eqs. (32) and (33), we omitted the subscript $p_r$ on $p$.

It follows that the clustering factors associated with the levels of the hierarchy additively contribute to the total variance. The expansion of $\sigma^2$ in terms of hierarchy levels might be useful for estimating $g_i$ from empirical data. Concerning IC manufacture, formulae (32) and (33) make it possible to develop the hierarchical analysis of variance for process-induced faults by analogy with that for product variables [27].

Even if an analytical formula is available for $G(z)$, the calculation of $P(k)$ from Eq. (28) is likely to be a time-consuming procedure, so it would be wise to perform numerical differentiation with a computer; an example is given in Table 1.

Table 1. Examples of distributions for different levels of modeling, with $n = 10^6$ and $p = 5 \cdot 10^{-7}$

| $k$ | Level 0 (Poisson model, $g_1 = g_2 = 0$) | Level 1 ($g_1 = 10^{-6}, g_2 = 0$) | Level 2 ($g_1 = g_2 = 10^{-6}$) |
|---|---|---|---|
| 0 | 0.6065 | 0.7071 | 0.7685 |
| 1 | 0.3033 | 0.1768 | 0.1135 |
| 2 | 0.0758 | 0.0663 | 0.0535 |
| 3 | 0.0126 | 0.0276 | 0.0282 |
| 4 | 0.0016 | 0.0121 | 0.0155 |
| 5 | 0.0002 | 0.0054 | 0.0088 |
| 6 | 0.0000 | 0.0025 | 0.0050 |
| 7 | 0.0000 | 0.0012 | 0.0029 |
| 8 | 0.0000 | 0.0005 | 0.0017 |
| 9 | 0.0000 | 0.0003 | 0.0010 |
| 10 | 0.0000 | 0.0001 | 0.0006 |
| 11 | 0.0000 | 0.0001 | 0.0003 |
| 12 | 0.0000 | 0.0000 | 0.0002 |
| 13 | 0.0000 | 0.0000 | 0.0001 |
| 14 | 0.0000 | 0.0000 | 0.0001 |
| 15 | 0.0000 | 0.0000 | 0.0000 |

Table 1 indicates that hierarchical clustering models predict a higher yield than simple Poisson models. At the same time, the former distributions show longer tails.

### 3.3 Correlation Characteristics in Multilevel Clustering Fault Model

Let us assume that the clustering factor $g_1$ describes the nonuniformity of the fault level from cluster to cluster (block) on a wafer; $g_2$, from wafer to wafer within a batch; $g_3$, from batch to batch etc.

From clustering nature of fault formation it follows that the numbers of these faults in closely spaced regions should correlate between each other stronger than in more distant ones. Such correlations can be described by the so-called interclass correlation coefficient introduced by Ronald Fisher for genetics problems.

Let us consider the stream of independent batches corresponding to the three level model. The interclass correlation coefficient reaches its peak for regions within a cluster (block):

$$\rho_1 = \frac{n(g_1 + g_2 + g_3)}{1 + n(g_1 + g_2 + g_3)}. \tag{34}$$

As the complexity $n$ of a region under control increases, this correlation coefficient tends to unity (the corresponding difference from unity is caused by the Poisson fluctuations in the number of faults). If the regions under control belong to different clusters within the same wafer, the correlation coefficient is:

$$\rho_2 = \frac{n(g_2 + g_3)}{1 + n(g_1 + g_2 + g_3)}. \tag{35}$$

The difference in these regions resulting in decreasing correlation compared to unity is caused not only by the Poisson fluctuations but also by the nonuniformity of the fault level from cluster to cluster within a wafer. Finally, the correlation coefficient for regions located on different wafers within a batch is equal to

$$\rho_3 = \frac{ng_3}{1 + n(g_1 + g_2 + g_3)}. \tag{36}$$

In this example, it is assumed that different batches do not correlate between each other: $\rho_4 = 0$. Generalization to the general case of arbitrary number of hierarchical levels is obvious. The correlations resulting from multilevel hierarchical clustering models describe well the results of numerical Monte Carlo simulations and real data.

## 4. YIELD ANALYSIS

**4.1 Yield Hierarchy**

In the absence of redundancy the yield is measured by the probability of choosing a fault-free chip ($k = 0$); therefore, it is simply equal to the value of the generating function at $z = 0$. Thus, for chips of complexity $n$, the generating-function hierarchy can easily be associated with a yield hierarchy. With $\ln Y_0 = -np_0$, we have

$$Y_0 = \exp(-np_0). \tag{37}$$

For levels 1 and 2,

$$\ln Y_1 = -p_1 n \frac{\ln(1 + g_1 n)}{g_1 n}, \tag{38}$$

$$\ln Y_2 = -p_2 n \frac{\ln\left(1 + g_2 n \frac{\ln(1 + g_1 n)}{g_1 n}\right)}{g_2 n}. \tag{39}$$

In general,

$$\ln Y_r = -p_r n \cdot L_r, \tag{40}$$

where

$$L_{r+1} = \frac{\ln(1 + g_{r+1} n L_r)}{g_{r+1} n}, L_0 = 1. \tag{41}$$

Comparing these with Eqs. (26) and (27), we see that $L_r \equiv L_r(z = 0)$.

To avoid confusion, we did not introduce new letters.

Table 2 gives examples of yield hierarchies for different IC complexities. The data reflect the fact that the Poisson model without clustering badly underestimates the yield of high-complexity ICs. Also notice that clustering is stronger in a level-2 model, provided that $g_1 = g_2$.

Table 2. Yield as a function of IC complexity for different levels of modeling ($p = 10^{-7}$)

| IC complexity | 256 K | 1 M | 4 M | 16 M | 64 M |
|---|---|---|---|---|---|
| Level 0 (Poisson model, $g_1 = g_2 = 0$) | 0.9741 | 0.9005 | 0.6574 | 0.1868 | 0.0012 |
| Level 1 ($g_1 = 5 \cdot 10^{-7}, g_2 = 0$) | 0.9757 | 0.9192 | 0.7976 | 0.6390 | 0.4924 |
| Level 2 ($g_1 = g_2 = 5 \cdot 10^{-7}$) | 0.9770 | 0.9321 | 0.8596 | 0.7905 | 0.7388 |

**4.2 Yield as a Random Variable: Its Distribution**

Let us consider a block (i.e., a limited area) on a silicon wafer [25]. With $p_0$ denoting the fault density for the block, the yield is given by the Poisson formula

$$Y_0 = \exp(-np_0).$$

Further, $p_0$ itself varies randomly from block to block according to a beta distribution. (In the situation considered, the beta distribution can be approximated by a gamma distribution to a very high accuracy.)

Let us change from the random variable $p_0$ to the random variable $Y_0 = \exp(-np_0)$. The PDF of the latter is expressed as follows [21]:

$$P(Y_0) = \frac{(b_1/n)^{a_1}}{\Gamma(a_1)} (-\ln Y_0)^{a_1-1} \cdot Y_0^{(b_1/n-1)}. \tag{42}$$

In general,

$$Y_r = \exp(-p_r n \cdot L_r), \tag{43}$$

so that

$$P(Y_r) = \frac{1}{\Gamma(a_{r+1})} \left(\frac{b_{r+1}}{nL_r}\right)^{a_{r+1}} (-\ln Y_r)^{a_{r+1}-1} Y_r^{\left(\frac{b_{r+1}}{nL_r}-1\right)}, \tag{44}$$

$$P(Y_r) = \frac{1}{\Gamma\left(\frac{p_{r+1}}{g_{r+1}}\right)} \left(\frac{1}{g_{r+1} nL_r}\right)^{\frac{p_{r+1}}{g_{r+1}}} (-\ln Y_r)^{\frac{p_{r+1}}{g_{r+1}}-1} Y_r^{\left(\frac{1}{g_{r+1} nL_r}-1\right)}. \tag{45}$$

Here, $Y_r$ is the yield (average value) for level $r$. In passing to level $r+1$, one treats $Y_r$ as a random variable with the PDF $P(Y_r)$. This approach enables one to naturally describe the random variation of yield from block to block within a wafer, from wafer to wafer within a batch, and so on.

Figure 1 shows yield PDFs for different IC complexities. The parameter $g$ refers to fault clustering within a wafer, whereas $a$ and $b$ serve to allow for the nonuniformity of fault distribution over the wafers.

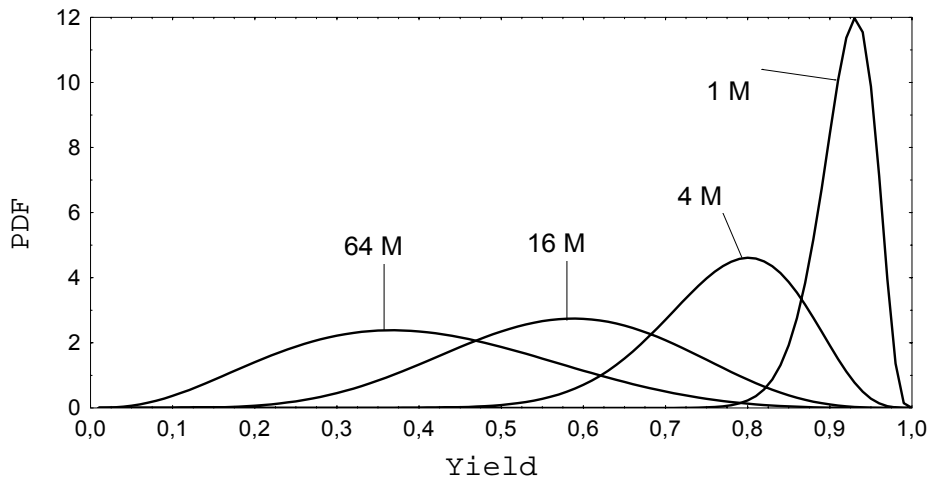

Fig. 1. Yield PDFs for different IC complexities ($g=3 \cdot 10^{-7}$, $a=5$, fault density $D=a/(a+b)=10^{-7}$)

## 5. CONCLUSIONS

We have developed a general approach to the construction of compound distributions for process-induced faults in IC manufacture. It quantifies the performance of a process in hierarchical form and helps one build general statistical models for fault control and yield management. With the aid of generating functions, main results can be written in compact analytical form. The strategy covers the negative binomial distribution regarding it as level-1 models.

The hierarchical approach to fault distribution offers an integrated picture of how fault density varies from area to area within a wafer, from wafer to wafer within a batch, and so on. The average fault density and the clustering factor can be estimated from an expansion of the variance in terms of hierarchy levels.

Yield hierarchies provide a new, general formalism for yield estimation and prediction in IC manufacture, treating the yield as a random variable. Particular cases are the Poisson distribution (level-0 model) and the negative binomial distribution (level-1 model).